\documentclass[onecolumn]{revtex4}
\usepackage{amsfonts}
\usepackage{amsmath}

\begin{document}

\title{New transformation of Wigner operator in phase space quantum mechanics for the
two-mode entangled case\thanks{{\small Work supported by the National Natural
Science Foundation of China under grant: 10775097 and 10874174 as well as the
President Foundation of Chinese Academy of Science}} }
\author{Hong-yi Fan$^{1,2}$ and Hong-chun Yuan$^{1}$\thanks{Corresponding author:
yuanhch@sjtu.edu.cn}\\$^{1}${\small Department of Physics, Shanghai
Jiao Tong University, \ Shanghai, 200030, China\\Corresponding
author: yuanhch@sjtu.edu.cn}\\$^{2}${\small Department of Material
Science and Engineering,}\\{\small University of Science and
Technology of China, Hefei, Anhui 230026, China}}

\begin{abstract}
As a natural extension of Fan's paper (arXiv: 0903.1769vl
[quant-ph]) by employing the formula of operators' Weyl ordering
expansion and the bipartite entangled state representation we find
new two-fold complex integration transformation about the Wigner
operator $\Delta \left(  \mu,\nu \right)  $ (in its entangled form)
in phase space quantum mechanics,
\begin{align*}
&  \int \frac{d^{2}\mu d^{2}\nu}{\pi^{2}}e^{\left(  \xi-\mu \right)  \left(
\eta^{\ast}-\nu^{\ast}\right)  -\left(  \eta-\nu \right)  \left(  \xi^{\ast
}-\mu^{\ast}\right)  }\Delta \left(  \mu,\nu \right) \\
&  =\delta \left(  \eta-a_{1}+a_{2}^{\dagger}\right)  \delta \left(  \eta^{\ast
}-a_{1}^{\dagger}+a_{2}\right)  \delta \left(  \xi-a_{1}-a_{2}^{\dagger
}\right)  \delta \left(  \xi^{\ast}-a_{1}^{\dagger}-a_{2}\right)  ,
\end{align*}
and its inverse transformation, where $a_{i},a_{i}^{\dagger}$ are bosonic
creation and annihilation operators, respectively. In this way, some operator
ordering problems regarding to $\left(  a_{1}^{\dagger}-a_{2}\right)  $ and
$\left(  a_{1}+a_{2}^{\dagger}\right)  $ can be solved and the contents of
phase space quantum mechanics can be enriched.



\end{abstract}
\maketitle

\section{Introduction}

\bigskip \ Integration transformations are very useful in mathematical physics.
In the preceding paper \cite{r1} we have reported a new integration
transformation in $q-p$ phase space%
\begin{equation}
\iint_{-\infty}^{\infty}\frac{dpdq}{\pi}e^{2i\left(  p-x\right)  \left(
q-y\right)  }h(p,q)=f\left(  x,y\right)  , \label{a1}%
\end{equation}
which is invertible%
\begin{equation}
\iint_{-\infty}^{\infty}\frac{dxdy}{\pi}e^{-2i(p-x)(q-y)}f(x,y)=h(p,q),
\label{a2}%
\end{equation}
and proved that this transformation obeys Parseval theorem. By virtue of the
formula of operators' Weyl ordering expansion \cite{r2,r3}, we find the
following new two-fold $q$-number integration transformation about the Wigner
operator $\Delta \left(  q^{\prime},p^{\prime}\right)  $ \cite{r4,r5,r6} in
phase space quantum mechanics \cite{r7},
\begin{equation}
\iint_{-\infty}^{\infty}\frac{\mathtt{d}p^{\prime}\mathtt{d}q^{\prime}}{\pi
}\Delta \left(  q^{\prime},p^{\prime}\right)  e^{-2i\left(  p-p^{\prime
}\right)  \left(  q-q^{\prime}\right)  }=\delta \left(  p-P\right)
\delta \left(  q-Q\right)  , \label{a3}%
\end{equation}
and its inverse%
\begin{equation}
\iint_{-\infty}^{\infty}\mathtt{d}q\mathtt{d}p\delta \left(  p-P\right)
\delta \left(  q-Q\right)  e^{2i\left(  p-p^{\prime}\right)  \left(
q-q^{\prime}\right)  }=\Delta \left(  q^{\prime},p^{\prime}\right)  ,
\label{a4}%
\end{equation}
where $Q,$ $P$ are the coordinate and momentum operators, respectively.
$\left[  Q,P\right]  =i\hbar.$ These two equations can be applied to studying
mutual converting formulas among $Q-P$ ordering, $P-Q$ ordering and Weyl
ordering of operators.

In the present paper, we extend Eqs.(\ref{a3}) and (\ref{a4}) to the two-mode
entangled case and construct the two-fold complex integration transformation
about the Wigner operator $\Delta \left(  \mu,\nu \right)  $ (in its entangled
form, $\mu,\nu$ are complex) in phase space quantum mechanics. By mentioning
two-mode entangled case, we naturally think of two mutually conjugate
entangled state representations $\left \vert \xi \right \rangle $ and $\left \vert
\eta \right \rangle $ (see Sect. 2 below). In Sect. 3, after deriving the Weyl
ordering of $\left \vert \eta \right \rangle \left \langle \xi \right \vert $ by
virtue of the formula of operator's Weyl ordering expansion, we give the Weyl
ordering form of

$\delta \left(  \nu-a_{1}+a_{2}^{\dagger}\right)  $ $\delta \left(  \nu^{\ast
}-a_{1}^{\dagger}+a_{2}\right)  $ $\delta \left(  \mu-a_{1}-a_{2}^{\dagger
}\right)  $ $\delta \left(  \mu^{\ast}-a_{1}^{\dagger}-a_{2}\right)  $, where
$a_{i},a_{i}^{\dagger}$ are the bosonic creation and annihilation operators
satisfying $\left[  a_{i},a_{j}^{\dagger}\right]  =\delta_{ij}$. Then in Sect.
4, we find a new two-fold complex integration transformation relating the
Wigner operator $\Delta \left(  \mu,\nu \right)  $ and the operator
$\delta \left(  \nu-a_{1}+a_{2}^{\dagger}\right)  $ $\delta \left(  \nu^{\ast
}-a_{1}^{\dagger}+a_{2}\right)  $ $\delta \left(  \mu-a_{1}-a_{2}^{\dagger
}\right)  $ $\delta \left(  \mu^{\ast}-a_{1}^{\dagger}-a_{2}\right)  .$ In
Sect. 5 we tackle some operator ordering problems regarding to $\left(
a_{1}^{\dagger}-a_{2}\right)  $ and $\left(  a_{1}+a_{2}^{\dagger}\right)  $.
In this way, the contents of phase space quantum mechanics can be further enriched.

\section{Brief review of the two mutually conjugate entangled states}

The conception of quantum entanglement has been paid much attention because of
their wide uses in quantum communication and quantum computation. The original
concept of quantum entanglement pointed out by Einstein, Podolsky and Rosen
(EPR) in their famous paper \cite{r8} arguing the incompleteness of quantum
mechanics is for a bipartite system characteristic of continuous variables.
According to the original idea of EPR that two particles' relative coordinate
operator commutes with their total momentum operator, $\left[  Q_{1}%
-Q_{2},P_{1}+P_{2}\right]  =0$, we have set up the bipartite entangle state
representation of continuous variable in two-mode Fock space \cite{r9}, which
is the common eigenvector $\left \vert \eta \right \rangle $ of $Q_{1}-Q_{2}$ and
$P_{1}+P_{2}$, i.e.
\begin{equation}
\left \vert \eta \right \rangle =\exp \left(  -\frac{1}{2}\left \vert
\eta \right \vert ^{2}+\eta a_{1}^{\dagger}-\eta^{\ast}a_{2}^{^{\dagger}}%
+a_{1}^{^{\dagger}}a_{2}^{^{\dagger}}\right)  \left \vert 00\right \rangle ,
\label{b1}%
\end{equation}
where $\eta \equiv \eta_{1}+i\eta_{2}$ is a complex number, $\left \vert
00\right \rangle $ is the two-mode vacuum state. Using $\left[  a_{i}%
,a_{j}^{\dagger}\right]  =\delta_{ij}$, it can be shown that $\left \vert
\eta \right \rangle $ obeys the eigenvector equations
\begin{equation}
\left(  a_{1}-a_{2}^{^{\dagger}}\right)  \left \vert \eta \right \rangle
=\eta \left \vert \eta \right \rangle ,\text{ }\left(  a_{2}-a_{1}^{^{\dagger}%
}\right)  \left \vert \eta \right \rangle =-\eta^{\ast}\left \vert \eta
\right \rangle .\text{ } \label{b2}%
\end{equation}
$\eta^{\prime}s$ real and imaginary part are respectively the eigenvalue of
$Q_{1}-Q_{2}$ and $P_{1}+P_{2}$,
\begin{equation}
\left(  Q_{1}-Q_{2}\right)  \left \vert \eta \right \rangle =\sqrt{2}\eta
_{1}\left \vert \eta \right \rangle ,\text{ }\left(  P_{1}+P_{2}\right)
\left \vert \eta \right \rangle =\sqrt{2}\eta_{2}\left \vert \eta \right \rangle .
\label{b4}%
\end{equation}
$\left \vert \eta \right \rangle $ is proved to be complete relation and
orthonormal property%
\begin{equation}
\int \frac{d^{2}\eta}{\pi}\left \vert \eta \right \rangle \left \langle
\eta \right \vert =1 \label{b6-1}%
\end{equation}
and%
\begin{equation}
\text{ }\left \langle \eta^{\prime}\right \vert \left.  \eta \right \rangle
=\pi \delta \left(  \eta^{\prime}-\eta \right)  \delta \left(  \eta^{\prime \ast
}-\eta^{\ast}\right)  \equiv \pi \delta^{\left(  2\right)  }\left(  \eta
^{\prime}-\eta \right)  . \label{b6-2}%
\end{equation}
On the other hand, $\left \vert \eta \right \rangle ^{\prime}$s conjugate state,
which is the common eigenvector of two particles' center-of-mass coordinate
and relative momentum, is%
\begin{equation}
\left \vert \xi \right \rangle =\exp \left(  -\frac{|\xi|^{2}}{2}+\xi
a_{1}^{\dagger}+\xi^{\ast}a_{2}^{\dagger}-a_{1}^{\dagger}a_{2}^{\dagger
}\right)  \left \vert 00\right \rangle , \label{b7}%
\end{equation}
with $\xi=\xi_{1}+i\xi_{2}$. $\left \vert \xi \right \rangle $ obeys
\begin{equation}
\left(  a_{1}+a_{2}^{\dagger}\right)  \left \vert \xi \right \rangle
=\xi \left \vert \xi \right \rangle ,\text{ \ }\left(  a_{1}^{\dagger}%
+a_{2}\right)  \left \vert \xi \right \rangle =\xi^{\ast}\left \vert
\xi \right \rangle , \label{b8}%
\end{equation}
and%
\begin{equation}
\left(  Q_{1}+Q_{2}\right)  \left \vert \xi \right \rangle =\sqrt{2}\xi
_{1}\left \vert \xi \right \rangle ,\text{ }\left(  P_{1}-P_{2}\right)
\left \vert \xi \right \rangle =\sqrt{2}\xi_{2}\left \vert \xi \right \rangle .
\label{b9}%
\end{equation}
$\left \vert \xi \right \rangle $ is completeness and orthonormal too,%
\begin{equation}
\int \frac{d^{2}\xi}{\pi}\left \vert \xi \right \rangle \left \langle
\xi \right \vert =1 \label{b10-1}%
\end{equation}
and%
\begin{equation}
\left \langle \xi \right.  \left \vert \xi^{\prime}\right \rangle =\pi
\delta \left(  \xi-\xi^{\prime}\right)  \delta \left(  \xi^{\ast}-\xi
^{\prime \ast}\right)  . \label{b10-2}%
\end{equation}
It is found that $\left \vert \xi \right \rangle $ and $\left \vert \eta
\right \rangle $ are of equal importance because
\begin{equation}
\left \langle \eta \right \vert \left.  \xi \right \rangle =\frac{1}{2}\exp \left[
\left(  \eta^{\ast}\xi-\xi^{\ast}\eta \right)  /2\right]  , \label{b11}%
\end{equation}
which is just a complex form of the Fourier transformation, since $\left(
\eta^{\ast}\xi-\xi^{\ast}\eta \right)  $ is pure imaginary, $\left \langle
\eta \right \vert \left.  \xi \right \rangle $ is a Fourier transform kernel.

\section{The Weyl ordering form of $\delta \left(  \nu-a_{1}+a_{2}^{\dagger
}\right)  $ $\delta \left(  \nu^{\ast}-a_{1}^{\dagger}+a_{2}\right)  $
$\delta \left(  \mu-a_{1}-a_{2}^{\dagger}\right)  $ $\delta \left(  \mu^{\ast
}-a_{1}^{\dagger}-a_{2}\right)  $}

Using the complete relation Eqs.(\ref{b6-1}) and (\ref{b10-1}), as well as
Eqs.(\ref{b2}), (\ref{b8}) and (\ref{b11}), we have%
\begin{align}
&  \delta \left(  \nu-a_{1}+a_{2}^{\dagger}\right)  \delta \left(  \nu^{\ast
}-a_{1}^{\dagger}+a_{2}\right)  \delta \left(  \mu-a_{1}-a_{2}^{\dagger
}\right)  \delta \left(  \mu^{\ast}-a_{1}^{\dagger}-a_{2}\right) \nonumber \\
&  =\int \frac{d^{2}\eta d^{2}\xi}{\pi^{2}}\delta^{\left(  2\right)  }\left(
\nu-a_{1}+a_{2}^{\dagger}\right)  \left \vert \eta \right \rangle \left \langle
\eta \right \vert \left.  \xi \right \rangle \left \langle \xi \right \vert
\delta^{\left(  2\right)  }\left(  \mu-a_{1}-a_{2}^{\dagger}\right)
\nonumber \\
&  =\frac{1}{2}\int \frac{d^{2}\eta d^{2}\xi}{\pi^{2}}\left \vert \eta
\right \rangle \left \langle \xi \right \vert e^{\left(  \eta^{\ast}\xi-\eta
\xi^{\ast}\right)  /2}\delta^{\left(  2\right)  }\left(  \nu-\eta \right)
\delta^{\left(  2\right)  }\left(  \mu-\xi \right) \nonumber \\
&  =\frac{1}{2}\left \vert \eta \right \rangle _{\eta=\nu}\left \langle
\xi \right \vert _{\xi=\mu}e^{\left(  \nu^{\ast}\mu-\nu \mu^{\ast}\right)  /2}.
\label{c1}%
\end{align}
where we have simplified the notation of the product of two Delta functions
\begin{align}
&  \delta^{\left(  2\right)  }\left(  \nu-a_{1}+a_{2}^{\dagger}\right)
\delta^{\left(  2\right)  }\left(  \mu-a_{1}-a_{2}^{\dagger}\right)
\nonumber \\
&  \equiv \delta \left(  \nu-a_{1}+a_{2}^{\dagger}\right)  \delta \left(
\nu^{\ast}-a_{1}^{\dagger}+a_{2}\right)  \delta \left(  \mu-a_{1}%
-a_{2}^{\dagger}\right)  \delta \left(  \mu^{\ast}-a_{1}^{\dagger}%
-a_{2}\right)  . \label{c1-1}%
\end{align}
From Eq.(\ref{c1}) we see that once the Weyl ordering of $\left \vert
\eta \right \rangle \left \langle \xi \right \vert $ is known, the Weyl ordering of
$\delta \left(  \nu-a_{1}+a_{2}^{\dagger}\right)  $ $\delta \left(  \nu^{\ast
}-a_{1}^{\dagger}+a_{2}\right)  $ $\delta \left(  \mu-a_{1}-a_{2}^{\dagger
}\right)  $ $\delta \left(  \mu^{\ast}-a_{1}^{\dagger}-a_{2}\right)  $ is
obtained. For this purpose, we recall the Weyl ordered expansion formula of
two-mode operators\cite{r2,r3}%
\begin{equation}
\rho=4\int \frac{d^{2}\beta_{1}d^{2}\beta_{2}}{\pi^{2}}%
\genfrac{}{}{0pt}{}{:}{:}%
\left \langle -\beta_{1},-\beta_{2}\right \vert \rho \left \vert \beta_{1}%
,\beta_{2}\right \rangle \exp \left[  2\sum \limits_{k=1}^{2}\left(  \beta
_{k}^{\ast}a_{k}-a_{k}^{\dagger}\beta_{k}+a_{k}^{\dagger}a_{k}\right)
\right]
\genfrac{}{}{0pt}{}{:}{:}
\label{c2}%
\end{equation}
where $%
\genfrac{}{}{0pt}{}{:}{:}%
\genfrac{}{}{0pt}{}{:}{:}%
$ denotes the Weyl ordering, within which $a_{i},a_{i}^{\dagger}$ can be
permuted, $\left \vert \beta_{1},\beta_{2}\right \rangle \equiv \left \vert
\beta_{1}\right \rangle \left \vert \beta_{2}\right \rangle $ is a two-mode
coherent state, $\left \vert \beta_{i}\right \rangle =\exp[-\frac{1}{2}%
|\beta_{i}|^{2}+\beta_{i}a_{i}^{\dagger}]\left \vert 0_{i}\right \rangle
$\cite{r10}$.$ For the operator $\left \vert \eta \right \rangle \left \langle
\xi \right \vert ,$ using Eq.(\ref{c2}) and the overlaps%
\begin{equation}
\left \langle \xi \right \vert \left.  \beta_{1},\beta_{2}\right \rangle
=\exp \left(  -\frac{1}{2}|\xi|^{2}+\xi^{\ast}\beta_{1}+\xi \beta_{2}-\beta
_{1}\beta_{2}-\frac{1}{2}|\beta_{1}|^{2}-\frac{1}{2}|\beta_{2}|^{2}\right)  ,
\label{c3}%
\end{equation}
and%
\begin{equation}
\left \langle -\beta_{1},-\beta_{2}\right \vert \left.  \eta \right \rangle
=\exp \left(  -\frac{1}{2}|\eta|^{2}-\eta \beta_{1}^{\ast}+\eta^{\ast}\beta
_{2}^{\ast}+\beta_{1}^{\ast}\beta_{2}^{\ast}-\frac{1}{2}|\beta_{1}|^{2}%
-\frac{1}{2}|\beta_{2}|^{2}\right)  , \label{c4}%
\end{equation}
we can derive the Weyl ordering of $\left \vert \eta \right \rangle \left \langle
\xi \right \vert $ in the way%

\begin{align}
\left \vert \eta \right \rangle \left \langle \xi \right \vert  &  =4\int \frac
{d^{2}\beta_{1}d^{2}\beta_{2}}{\pi^{2}}%
\genfrac{}{}{0pt}{}{:}{:}%
\left \langle -\beta_{1},-\beta_{2}\right \vert \left.  \eta \right \rangle
\left \langle \xi \right \vert \left.  \beta_{1},\beta_{2}\right \rangle
\exp \left[  2\sum \limits_{k=1}^{2}\left(  \beta_{k}^{\ast}a_{k}-a_{k}%
^{\dagger}\beta_{k}+a_{k}^{\dagger}a_{k}\right)  \right]
\genfrac{}{}{0pt}{}{:}{:}%
\nonumber \\
&  =4\int \frac{d^{2}\beta_{1}d^{2}\beta_{2}}{\pi^{2}}%
\genfrac{}{}{0pt}{}{:}{:}%
\exp \left[  -\frac{|\eta|^{2}}{2}-\eta \beta_{1}^{\ast}+\eta^{\ast}\beta
_{2}^{\ast}+\beta_{1}^{\ast}\beta_{2}^{\ast}-|\beta_{1}|^{2}-|\beta_{2}%
|^{2}\right. \nonumber \\
&  \left.  -\frac{|\xi|^{2}}{2}+\xi^{\ast}\beta_{1}+\xi \beta_{2}-\beta
_{1}\beta_{2}+2\sum \limits_{k=1}^{2}\left(  \beta_{k}^{\ast}a_{k}%
-a_{k}^{\dagger}\beta_{k}+a_{k}^{\dagger}a_{k}\right)  \right]
\genfrac{}{}{0pt}{}{:}{:}%
\nonumber \\
&  =4e^{-\frac{|\eta|^{2}}{2}-\frac{|\xi|^{2}}{2}}\int \frac{d^{2}\beta_{2}%
}{\pi}%
\genfrac{}{}{0pt}{}{:}{:}%
\exp \left[  -2|\beta_{2}|^{2}+\beta_{2}\left(  \xi-2a_{2}^{\dagger}%
-2a_{1}+\eta \right)  \right. \nonumber \\
&  \left.  +\beta_{2}^{\ast}\left(  \eta^{\ast}+2a_{2}+\xi^{\ast}%
-2a_{1}^{\dagger}\right)  +\left(  \xi^{\ast}-2a_{1}^{\dagger}\right)  \left(
2a_{1}-\eta \right)  +2\sum_{k=1}^{2}a_{k}^{\dagger}a_{k}\right]
\genfrac{}{}{0pt}{}{:}{:}%
\nonumber \\
&  =2%
\genfrac{}{}{0pt}{}{:}{:}%
\exp \left[  \frac{1}{2}\left(  \xi \eta^{\ast}-\eta \xi^{\ast}\right)
+\xi \left(  a_{2}-a_{1}^{\dagger}\right)  +\eta \left(  a_{2}+a_{1}^{\dagger
}\right)  \right. \nonumber \\
&  \left.  -\eta^{\ast}\left(  a_{1}+a_{2}^{\dagger}\right)  +\xi^{\ast
}\left(  a_{1}-a_{2}^{\dagger}\right)  +2a_{2}^{\dagger}a_{1}^{\dagger}%
-2a_{2}a_{1}\right]
\genfrac{}{}{0pt}{}{:}{:}%
, \label{c5}%
\end{align}
where we have used the integral formula%
\begin{equation}
\int \frac{d^{2}z}{\pi}\exp \left(  \zeta \left \vert z\right \vert ^{2}+\xi z+\eta
z^{\ast}\right)  =-\frac{1}{\zeta}\exp \left[  -\frac{\xi \eta}{\zeta}\right]
,\operatorname{Re}\left(  \xi \right)  <0 \label{c6}%
\end{equation}
It then follows from Eq.(\ref{c5}) that%
\begin{equation}
\frac{1}{2}\left \vert \eta \right \rangle \left \langle \xi \right \vert e^{\left(
\eta^{\ast}\xi-\eta \xi^{\ast}\right)  /2}=%
\genfrac{}{}{0pt}{}{:}{:}%
\exp[\left(  \xi-a_{1}-a_{2}^{\dagger}\right)  \left(  \eta^{\ast}-a_{1}%
^{\dag}+a_{2}\right)  -\left(  \eta-a_{1}+a_{2}^{\dagger}\right)  \left(
\xi^{\ast}-a_{1}^{\dag}-a_{2}\right)  ]%
\genfrac{}{}{0pt}{}{:}{:}%
. \label{c7}%
\end{equation}
Substituting Eq.(\ref{c7}) into Eq.(\ref{c1}) we have the Weyl ordered form
\begin{align}
&  \delta \left(  \nu-a_{1}+a_{2}^{\dagger}\right)  \delta \left(  \nu^{\ast
}-a_{1}^{\dagger}+a_{2}\right)  \delta \left(  \mu-a_{1}-a_{2}^{\dagger
}\right)  \delta \left(  \mu^{\ast}-a_{1}^{\dagger}-a_{2}\right) \nonumber \\
&  =%
\genfrac{}{}{0pt}{}{:}{:}%
\exp[\left(  \mu-a_{1}-a_{2}^{\dagger}\right)  \left(  \nu^{\ast}-a_{1}^{\dag
}+a_{2}\right)  -\left(  \nu-a_{1}+a_{2}^{\dagger}\right)  \left(  \mu^{\ast
}-a_{1}^{\dag}-a_{2}\right)  ]%
\genfrac{}{}{0pt}{}{:}{:}%
, \label{c8}%
\end{align}
\textbf{\ }Noting
\begin{equation}
\left[  a_{1}^{\dagger}-a_{2},a_{1}+a_{2}^{\dagger}\right]  =-2, \label{c9}%
\end{equation}
we should consider another ordering other than Eq.(\ref{c8}), i.e.,
\begin{align}
&  \delta \left(  \mu-a_{1}-a_{2}^{\dagger}\right)  \delta \left(  \mu^{\ast
}-a_{1}^{\dagger}-a_{2}\right)  \delta \left(  \nu-a_{1}+a_{2}^{\dagger
}\right)  \delta \left(  \nu^{\ast}-a_{1}^{\dagger}+a_{2}\right) \nonumber \\
&  =\frac{1}{2}\int \frac{d^{2}\eta d^{2}\xi}{\pi^{2}}\left \vert \xi
\right \rangle \left \langle \eta \right \vert e^{\left(  \eta \xi^{\ast}%
-\eta^{\ast}\xi \right)  /2}\delta^{\left(  2\right)  }\left(  \nu-\eta \right)
\delta^{\left(  2\right)  }\left(  \mu-\xi \right)  . \label{c10}%
\end{align}
Similar in the way of deriving Eq.(\ref{c7}), we deduce%

\begin{equation}
\frac{1}{2}\left \vert \xi \right \rangle \left \langle \eta \right \vert e^{\left(
\eta \xi^{\ast}-\eta^{\ast}\xi \right)  /2}=%
\genfrac{}{}{0pt}{}{:}{:}%
\exp \left[  \left(  \eta-\allowbreak a_{1}+a_{2}^{\dagger}\right)  \left(
\xi^{\ast}-a_{1}^{\dagger}-a_{2}\right)  -\left(  \xi-a_{1}-a_{2}^{\dagger
}\right)  \allowbreak \left(  \eta^{\ast}-a_{1}^{\dagger}+a_{2}\right)
\allowbreak \right]
\genfrac{}{}{0pt}{}{:}{:}%
.\label{c11}%
\end{equation}%
\[%
\genfrac{}{}{0pt}{}{:}{:}%
\exp \left[  \left(  \eta-\allowbreak a_{1}+a_{2}^{\dagger}\right)  \left(
\xi^{\ast}-a_{1}^{\dagger}-a_{2}\right)  -\left(  \allowbreak \eta^{\ast}%
-a_{1}^{\dagger}+a_{2}\right)  \left(  \xi-\allowbreak a_{2}^{\dagger}%
-a_{1}\right)  \right]
\genfrac{}{}{0pt}{}{:}{:}%
\]
It then follows
\begin{align}
&  \delta \left(  \mu-a_{1}-a_{2}^{\dagger}\right)  \delta \left(  \mu^{\ast
}-a_{1}^{\dagger}-a_{2}\right)  \delta \left(  \nu-a_{1}+a_{2}^{\dagger
}\right)  \delta \left(  \nu^{\ast}-a_{1}^{\dagger}+a_{2}\right)  \nonumber \\
&  =%
\genfrac{}{}{0pt}{}{:}{:}%
\exp \left[  \left(  \nu-\allowbreak a_{1}+a_{2}^{\dagger}\right)  \left(
\mu^{\ast}-a_{1}^{\dagger}-a_{2}\right)  -\left(  \mu-a_{1}-a_{2}^{\dagger
}\right)  \allowbreak \left(  \nu^{\ast}-a_{1}^{\dagger}+a_{2}\right)
\allowbreak \right]
\genfrac{}{}{0pt}{}{:}{:}%
.\label{c12}%
\end{align}

\section{The mutual integration transformation between $\delta \left(
\nu-a_{1}+a_{2}^{\dagger}\right)  $ $\delta \left(  \nu^{\ast}-a_{1}^{\dagger
}+a_{2}\right)  $ $\delta \left(  \mu-a_{1}-a_{2}^{\dagger}\right)  $
$\delta \left(  \mu^{\ast}-a_{1}^{\dagger}-a_{2}\right)  $ and the Wigner
operator}

In Ref. \cite{r11} for correlated two-body systems, we have successfully
derived the Wigner operator in entangled form, expressed in the entangled
state $\left \langle \eta \right \vert $ representation as
\begin{equation}
\Delta \left(  \mu,\nu \right)  =\int \frac{d^{2}\eta}{\pi^{3}}\left \vert
\nu-\eta \right \rangle \left \langle \nu+\eta \right \vert e^{\eta \mu^{\ast}%
-\eta^{\ast}\mu}, \label{e1}%
\end{equation}
$\Delta \left(  \mu,\nu \right)  $ plays the role of establishing
the relationship between $\eta-\xi$ phase space function and its
Weyl-Wigner quantum correspondence operator. Using the similar
method to deriving Eq.(\ref{c5}), we find that the Weyl ordering of
operator $\left \vert \nu -\eta \right \rangle \left \langle
\nu+\eta \right \vert $ is
\begin{align}
\left \vert \nu-\eta \right \rangle \left \langle \nu+\eta \right \vert  &  =%
\genfrac{}{}{0pt}{}{:}{:}%
\delta \left(  \nu-a_{1}+a_{2}^{\dagger}\right)  \delta \left(  \nu^{\ast}%
-a_{1}^{\dagger}+a_{2}\right) \nonumber \\
&  \times \exp \left[  \eta \left(  \nu^{\ast}-2a_{1}^{\dagger}\right)
-\eta^{\ast}\left(  \nu-2a_{1}\right)  -2\left(  a_{1}^{\dagger}-\nu^{\ast
}\right)  \left(  a_{1}-\nu \right)  +2a_{2}^{\dagger}a_{2}\right]
\genfrac{}{}{0pt}{}{:}{:}%
. \label{e2}%
\end{align}
Substituting (\ref{e2}) into Eq.(\ref{e1}) and performing the integration over
$d^{2}\eta,$ the result is%
\begin{align}
\Delta \left(  \mu,\nu \right)   &  =%
\genfrac{}{}{0pt}{}{:}{:}%
\delta \left(  \mu-a_{1}-a_{2}^{\dagger}\right)  \delta \left(  \mu^{\ast}%
-a_{1}^{\dagger}-a_{2}\right)  \delta \left(  \nu-a_{1}+a_{2}^{\dagger}\right)
\delta \left(  \nu^{\ast}-a_{1}^{\dagger}+a_{2}\right)
\genfrac{}{}{0pt}{}{:}{:}%
\nonumber \\
&  =%
\genfrac{}{}{0pt}{}{:}{:}%
\delta^{\left(  2\right)  }\left(  \mu-a_{1}-a_{2}^{\dagger}\right)
\delta^{\left(  2\right)  }\left(  \nu-a_{1}+a_{2}^{\dagger}\right)
\genfrac{}{}{0pt}{}{:}{:}%
, \label{e3}%
\end{align}
which is also a neat Dirac's $\delta$-operator function within the Weyl
ordering symbol.

As a result of Eqs.(\ref{c8}) and (\ref{e3}), we have
\begin{align}
&  \delta \left(  \eta-a_{1}+a_{2}^{\dagger}\right)  \delta \left(  \eta^{\ast
}-a_{1}^{\dagger}+a_{2}\right)  \delta \left(  \xi-a_{1}-a_{2}^{\dagger
}\right)  \delta \left(  \xi^{\ast}-a_{1}^{\dagger}-a_{2}\right) \nonumber \\
&  =%
\genfrac{}{}{0pt}{}{:}{:}%
\exp[\left(  \xi-a_{1}-a_{2}^{\dagger}\right)  \left(  \eta^{\ast}-a_{1}%
^{\dag}+a_{2}\right)  -\left(  \eta-a_{1}+a_{2}^{\dagger}\right)  \left(
\xi^{\ast}-a_{1}^{\dag}-a_{2}\right)  ]%
\genfrac{}{}{0pt}{}{:}{:}%
\nonumber \\
&  =\int \frac{d^{2}\mu d^{2}\nu}{\pi^{2}}%
\genfrac{}{}{0pt}{}{:}{:}%
e^{\left(  \xi-\mu \right)  \left(  \eta^{\ast}-\nu^{\ast}\right)  -\left(
\eta-\nu \right)  \left(  \xi^{\ast}-\mu^{\ast}\right)  }\delta^{\left(
2\right)  }\left(  \mu-a_{1}-a_{2}^{\dagger}\right)  \delta^{\left(  2\right)
}\left(  \nu-a_{1}+a_{2}^{\dagger}\right)
\genfrac{}{}{0pt}{}{:}{:}%
\nonumber \\
&  =\int \frac{d^{2}\mu d^{2}\nu}{\pi^{2}}e^{\left(  \xi-\mu \right)  \left(
\eta^{\ast}-\nu^{\ast}\right)  -\left(  \eta-\nu \right)  \left(  \xi^{\ast
}-\mu^{\ast}\right)  }\Delta \left(  \mu,\nu \right)  . \label{e4}%
\end{align}
This is the mutual transformation between the entangled form of Wigner
operator $\Delta \left(  \mu,\nu \right)  $ and $\delta \left(  \mu-a_{1}%
-a_{2}^{\dagger}\right)  $ $\delta \left(  \mu^{\ast}-a_{1}^{\dagger}%
-a_{2}\right)  $ $\delta \left(  \nu-a_{1}+a_{2}^{\dagger}\right)  $
$\delta \left(  \nu^{\ast}-a_{1}^{\dagger}+a_{2}\right)  .$

Moveover, the reciprocal transformation of Eq. (\ref{e4}) is%
\begin{align}
&  \int \frac{d^{2}\xi d^{2}\eta}{\pi^{2}}\delta \left(  \eta-a_{1}%
+a_{2}^{\dagger}\right)  \delta \left(  \eta^{\ast}-a_{1}^{\dagger}%
+a_{2}\right)  \delta \left(  \xi-a_{1}-a_{2}^{\dagger}\right)  \delta \left(
\xi^{\ast}-a_{1}^{\dagger}-a_{2}\right)  e^{-\left(  \xi-\mu \right)  \left(
\eta^{\ast}-\nu^{\ast}\right)  +\left(  \eta-\nu \right)  \left(  \xi^{\ast
}-\mu^{\ast}\right)  }\nonumber \\
&  =\int \frac{d^{2}\xi d^{2}\eta}{\pi^{2}}\int \frac{d^{2}\mu^{\prime}d^{2}%
\nu^{\prime}}{\pi^{2}}\Delta \left(  \mu^{\prime},\nu^{\prime}\right)
e^{\left(  \xi-\mu^{\prime}\right)  \left(  \eta^{\ast}-\nu^{\prime \ast
}\right)  -\left(  \eta-\nu^{\prime}\right)  \left(  \xi^{\ast}-\mu
^{\prime \ast}\right)  }e^{-\left(  \xi-\mu \right)  \left(  \eta^{\ast}%
-\nu^{\ast}\right)  +\left(  \eta-\nu \right)  \left(  \xi^{\ast}-\mu^{\ast
}\right)  }\nonumber \\
&  =\int \frac{d^{2}\mu^{\prime}d^{2}\nu^{\prime}}{\pi^{2}}\Delta \left(
\mu^{\prime},\nu^{\prime}\right)  e^{\left(  -\nu^{\prime}\mu^{\prime \ast}%
+\mu^{\prime}\nu^{\prime \ast}-\mu \nu^{\ast}+\nu \mu^{\ast}\right)  }\int
\frac{d^{2}\xi d^{2}\eta}{\pi^{2}}e^{\xi \left(  \nu^{\ast}-\nu^{\prime \ast
}\right)  +\xi^{\ast}\left(  \nu^{\prime}-\nu \right)  }e^{\eta^{\ast}\left(
\mu-\mu^{\prime}\right)  +\eta \left(  \mu^{\prime \ast}-\mu^{\ast}\right)
}\nonumber \\
&  =\int \frac{d^{2}\mu^{\prime}d^{2}\nu^{\prime}}{\pi^{2}}\Delta \left(
\mu^{\prime},\nu^{\prime}\right)  e^{\left(  -\nu^{\prime}\mu^{\prime \ast}%
+\mu^{\prime}\nu^{\prime \ast}-\mu \nu^{\ast}+\nu \mu^{\ast}\right)  }%
\delta^{\left(  2\right)  }\left(  \nu^{\prime}-\nu \right)  \delta^{\left(
2\right)  }\delta \left(  \mu^{\prime}-\mu \right) \nonumber \\
&  =\Delta \left(  \mu,\nu \right)  . \label{e5}%
\end{align}
So Eqs.(\ref{e4}) and (\ref{e5}) are qualified to be new two-fold complex
integration transformations about the Wigner operator $\Delta \left(  \mu
,\nu \right)  $ (in its entangled form) in phase space quantum mechanics.

Further, multiplying both sides of (\ref{e5}) from the left by $\int
\frac{d^{2}\mu d^{2}\nu}{\pi^{2}}D\left(  \nu,\mu \right)  $ and considering
\begin{equation}
Q_{j}=\frac{a_{j}+a_{j}^{^{\dagger}}}{\sqrt{2}},\text{ }P_{j}=\frac
{a_{j}-a_{j}^{^{\dagger}}}{i\sqrt{2}}, \label{b3}%
\end{equation}
we obtain%
\begin{align}
&  \int \frac{d^{2}\mu d^{2}\nu}{\pi^{2}}D\left(  \nu,\mu \right)  \Delta \left(
\mu,\nu \right) \nonumber \\
&  =\int \frac{d^{2}\xi d^{2}\eta}{\pi^{2}}\delta^{\left(  2\right)  }\left(
\eta-a_{1}+a_{2}^{\dagger}\right)  \delta^{\left(  2\right)  }\left(
\xi-a_{1}-a_{2}^{\dagger}\right)  \int \frac{d^{2}\mu d^{2}\nu}{\pi^{2}%
}D\left(  \nu,\mu \right)  e^{-\left(  \xi-\mu \right)  \left(  \eta^{\ast}%
-\nu^{\ast}\right)  +\left(  \eta-\nu \right)  \left(  \xi^{\ast}-\mu^{\ast
}\right)  }\nonumber \\
&  =\int \frac{d^{2}\xi d^{2}\eta}{\pi^{2}}\delta^{\left(  2\right)  }\left(
\eta-a_{1}+a_{2}^{\dagger}\right)  \delta^{\left(  2\right)  }\left(
\xi-a_{1}-a_{2}^{\dagger}\right)  \mathcal{F}(\eta,\xi)\nonumber \\
&  =\int \frac{d^{2}\xi d^{2}\eta}{\pi^{2}}\delta \left(  \eta_{1}-\frac
{Q_{1}-Q_{2}}{\sqrt{2}}\right)  \delta \left(  \eta_{2}-\frac{P_{1}+P_{2}%
}{\sqrt{2}}\right)  \delta \left(  \xi_{1}-\frac{Q_{1}+Q_{2}}{\sqrt{2}}\right)
\delta \left(  \xi_{2}-\frac{P_{1}-P_{2}}{\sqrt{2}}\right)  \mathcal{F}%
(\eta,\xi)\nonumber \\
&  =\mathcal{F}(\frac{Q_{1}-Q_{2}}{\sqrt{2}},\frac{P_{1}+P_{2}}{\sqrt{2}%
},\frac{Q_{1}+Q_{2}}{\sqrt{2}},\frac{P_{1}-P_{2}}{\sqrt{2}}), \label{e7}%
\end{align}
where we have introduced%
\begin{equation}
\mathcal{F}(\eta,\xi)\equiv \int \frac{d^{2}\mu d^{2}\nu}{\pi^{2}}D\left(
\nu,\mu \right)  e^{\left(  \xi^{\ast}-\mu^{\ast}\right)  \left(  \eta
-\nu \right)  -\left(  \eta^{\ast}-\nu^{\ast}\right)  \left(  \xi-\mu \right)
}, \label{e8}%
\end{equation}
which seems a new interesting transformation. Due to%
\begin{align}
&  \int \frac{d^{2}\xi d^{2}\eta}{\pi^{2}}\exp \left[  \left(  \xi-\mu \right)
\left(  \eta^{\ast}-\nu^{\ast}\right)  -\left(  \eta-\nu \right)  \left(
\xi^{\ast}-\mu^{\ast}\right)  \right] \nonumber \\
&  =\int d^{2}\xi \delta \left(  \xi-\mu \right)  \delta \left(  \xi^{\ast}%
-\mu^{\ast}\right)  e^{\nu \left(  \xi^{\ast}-\mu^{\ast}\right)  -\nu^{\ast
}\left(  \xi-\mu \right)  }=1, \label{e6}%
\end{align}
$e^{\left(  \xi-\mu \right)  \left(  \eta^{\ast}-\nu^{\ast}\right)  -\left(
\eta-\nu \right)  \left(  \xi^{\ast}-\mu^{\ast}\right)  }$ can be considered a
basis function in $\xi-\eta$ phase space, or Eq. (\ref{e4}) can be looked as
an expansion of $D\left(  \nu,\mu \right)  $ in terms of $e^{\left(  \xi
-\mu \right)  \left(  \eta^{\ast}-\nu^{\ast}\right)  -\left(  \eta-\nu \right)
\left(  \xi^{\ast}-\mu^{\ast}\right)  },$ with the expansion coefficient being
$\mathcal{F}(\eta,\xi)$.

We can prove that the inverse transform of (\ref{e8}) is
\begin{equation}
\int \frac{d^{2}\xi d^{2}\eta}{\pi^{2}}e^{\left(  \xi-\mu \right)  \left(
\eta^{\ast}-\nu^{\ast}\right)  -\left(  \eta-\nu \right)  \left(  \xi^{\ast
}-\mu^{\ast}\right)  }\mathcal{F}(\eta,\xi)\equiv D\left(  \nu,\mu \right)
.\label{e9}%
\end{equation}
In fact, substituting Eq. (\ref{e9}) into Eq. (\ref{e8}) yields%
\begin{align}
&  \int \frac{d^{2}\xi^{\prime}d^{2}\eta^{\prime}}{\pi^{2}}\mathcal{F}%
(\eta^{\prime},\xi^{\prime})\int \frac{d^{2}\mu d^{2}\nu}{\pi^{2}}e^{\left(
\xi^{\prime}-\mu \right)  \left(  \eta^{\prime \ast}-\nu^{\ast}\right)  -\left(
\eta^{\prime}-\nu \right)  \left(  \xi^{\prime \ast}-\mu^{\ast}\right)  +\left(
\xi^{\ast}-\mu^{\ast}\right)  \left(  \eta-\nu \right)  -\left(  \eta^{\ast
}-\nu^{\ast}\right)  \left(  \xi-\mu \right)  }\nonumber \\
&  =\int \frac{d^{2}\xi^{\prime}d^{2}\eta^{\prime}}{\pi^{2}}\mathcal{F}%
(\eta^{\prime},\xi^{\prime})e^{\left(  \xi^{\prime}\eta^{\prime \ast}%
-\eta^{\prime}\xi^{\prime \ast}+\xi^{\ast}\eta-\eta^{\ast}\xi \right)  }%
\int \frac{d^{2}\mu d^{2}\nu}{\pi^{2}}e^{\left(  \eta^{\ast}-\eta^{\prime \ast
}\right)  \mu+\left(  \eta^{\prime}-\eta \right)  \mu^{\ast}}e^{\left(
\xi^{\prime \ast}-\xi^{\ast}\right)  \nu+\left(  \xi-\xi^{\prime}\right)
\nu^{\ast}}\nonumber \\
&  =\int d^{2}\xi^{\prime}d^{2}\eta^{\prime}\mathcal{F}(\eta^{\prime}%
,\xi^{\prime})e^{\left(  \xi^{\prime}\eta^{\prime \ast}-\eta^{\prime}%
\xi^{\prime \ast}+\xi^{\ast}\eta-\eta^{\ast}\xi \right)  }\delta^{\left(
2\right)  }\left(  \eta^{\prime}-\eta \right)  \delta^{\left(  2\right)
}\delta \left(  \xi-\xi^{\prime}\right)  \nonumber \\
&  =\mathcal{F}(\eta,\xi).\label{e10}%
\end{align}
Additionally, we can prove
\begin{align}
&  \int \frac{d^{2}\xi d^{2}\eta}{\pi^{2}}|\mathcal{F}(\eta,\xi)|^{2}%
\nonumber \\
&  =\int \frac{d^{2}\mu d^{2}\nu}{\pi^{2}}|D\left(  \nu,\mu \right)  |^{2}%
\int \frac{d^{2}\mu^{\prime}d^{2}\nu^{\prime}}{\pi^{2}}\exp \left[  \left(
\mu^{\ast}\nu-\nu^{\ast}\mu \right)  +\left(  \mu^{\prime}\nu^{\prime \ast}%
-\nu^{\prime}\mu^{\prime \ast}\right)  \right]  \nonumber \\
&  \times \int \frac{d^{2}\xi d^{2}\eta}{\pi^{2}}\exp \left[  \left(  \mu
^{\prime \ast}-\mu^{\ast}\right)  \eta+\left(  \mu-\mu^{\prime}\right)
\eta^{\ast}+(\nu^{\ast}-\nu^{\prime \ast})\xi+(\nu^{\prime}-\nu)\xi^{\ast
}\right]  \nonumber \\
&  =\int \frac{d^{2}\mu d^{2}\nu}{\pi^{2}}|D\left(  \nu,\mu \right)  |^{2}\int
d^{2}\mu^{\prime}d^{2}\nu^{\prime}\exp \left[  \left(  \mu^{\ast}\nu-\nu^{\ast
}\mu \right)  +\left(  \mu^{\prime}\nu^{\prime \ast}-\nu^{\prime}\mu^{\prime
\ast}\right)  \right]  \delta^{\left(  2\right)  }\left(  \mu-\mu^{\prime
}\right)  \delta^{\left(  2\right)  }\delta(\nu^{\prime}-\nu)\nonumber \\
&  =\int \frac{d^{2}\mu d^{2}\nu}{\pi^{2}}|D\left(  \nu,\mu \right)
|^{2},\label{e12}%
\end{align}
which shows that the transformations of Eqs.(\ref{e8}) and (\ref{e9}) obey the
Parseval-like theorem as well.

\section{Application}

By using the above transformation we can solve some operator ordering problems
arising from Eq.(\ref{c9}), for example, we want to put $\left(
a_{1}^{\dagger}-a_{2}\right)  ^{n}\left(  a_{1}+a_{2}^{\dagger}\right)  ^{m}$
into its Weyl ordered form. According to Eqs.(\ref{b6-1}), (\ref{b10-1}) and
(\ref{b11}), we have%
\begin{equation}
\frac{1}{2}\int \frac{d^{2}\eta d^{2}\xi}{\pi^{2}}\left \vert \eta \right \rangle
\left \langle \xi \right \vert e^{\left(  \eta^{\ast}\xi-\eta \xi^{\ast}\right)
/2}=1 \label{g1}%
\end{equation}
and
\begin{equation}
\frac{1}{2}\int \frac{d^{2}\eta d^{2}\xi}{\pi^{2}}\left \vert \xi \right \rangle
\left \langle \eta \right \vert e^{\left(  \eta \xi^{\ast}-\eta^{\ast}\xi \right)
/2}=1. \label{g2}%
\end{equation}
As a result of Eqs.(\ref{g1}) and (\ref{c7}), using the eigenvector equations
Eqs.(\ref{b2}) and (\ref{b8}), we obtain \bigskip%
\begin{align}
&  \left(  a_{1}^{\dagger}-a_{2}\right)  ^{n}\left(  a_{1}+a_{2}^{\dagger
}\right)  ^{m}\nonumber \\
&  =\frac{1}{2}\int \frac{d^{2}\eta d^{2}\xi}{\pi^{2}}\eta^{\ast n}\xi
^{m}\left \vert \eta \right \rangle \left \langle \xi \right \vert e^{\left(
\eta^{\ast}\xi-\eta \xi^{\ast}\right)  /2}\nonumber \\
&  =\int \frac{d^{2}\eta d^{2}\xi}{\pi^{2}}\eta^{\ast n}\xi^{m}%
\genfrac{}{}{0pt}{}{:}{:}%
\exp[\left(  \xi-a_{1}-a_{2}^{\dagger}\right)  \left(  \eta^{\ast}-a_{1}%
^{\dag}+a_{2}\right)  -\left(  \eta-a_{1}+a_{2}^{\dagger}\right)  \left(
\xi^{\ast}-a_{1}^{\dag}-a_{2}\right)  ]%
\genfrac{}{}{0pt}{}{:}{:}%
\nonumber \\
&  =\int \frac{d^{2}\eta d^{2}\xi}{\pi^{2}}\sum_{k=0}^{n}\sum_{l=0}^{m}%
\frac{i^{n+m}m!n!}{\left(  n-k\right)  !k!\left(  m-l\right)  !l!}\eta
_{1}^{n-k}\left(  -\eta_{2}\right)  ^{n}\xi_{1}^{m-l}\xi_{2}^{m}\nonumber \\
&  \times%
\genfrac{}{}{0pt}{}{:}{:}%
\exp \left \{  \left[  -2i\eta_{2}\allowbreak+\left(  a_{1}+a_{2}-a_{2}%
^{\dagger}-a_{1}^{\dag}\right)  \right]  \left[  \xi_{1}-\frac{1}{2}\left(
a_{2}^{\dagger}+a_{1}^{\dag}+a_{1}+a_{2}\right)  \right]  \right. \nonumber \\
&  \left.  +\left[  2i\allowbreak \eta_{1}+i\left(  a_{2}^{\dagger}-\allowbreak
a_{1}^{\dag}-a_{1}+a_{2}\right)  \right]  \left[  \xi_{2}-\frac{1}{2}i\left(
a_{1}^{\dag}-a_{1}+\allowbreak a_{2}-a_{2}^{\dagger}\right)  \right]
\right \}
\genfrac{}{}{0pt}{}{:}{:}%
\nonumber \\
&  =-\sum_{k=0}^{n}\sum_{l=0}^{m}\frac{i^{n+m}m!n!}{\left(  n-k\right)
!k!\left(  m-l\right)  !l!}\nonumber \\
&  \times \int \frac{d\left(  -\eta_{2}\right)  d\xi_{1}}{\pi}\left(  -\eta
_{2}\right)  ^{n}\xi_{1}^{m-l}%
\genfrac{}{}{0pt}{}{:}{:}%
\exp \left \{  2i\left[  -\eta_{2}\allowbreak+\frac{1}{\sqrt{2}}\left(
P_{1}+P_{2}\right)  \right]  \left[  \xi_{1}-\frac{1}{\sqrt{2}}\left(
Q_{1}+Q_{2}\right)  \right]  \right \} \nonumber \\
&  \times \int \frac{d\eta_{1}d\xi_{2}}{\pi}\eta_{1}^{n-k}\xi_{2}^{m}%
\exp \left \{  2i\allowbreak \left[  \eta_{1}-\frac{1}{\sqrt{2}}\left(
Q_{1}-Q_{2}\right)  \right]  \left[  \xi_{2}-\frac{1}{\sqrt{2}}\left(
P_{1}-P_{2}\right)  \right]  \right \}
\genfrac{}{}{0pt}{}{:}{:}%
\nonumber \\
&  =-\left(  \frac{1}{2}\right)  ^{n+m}\sum_{k=0}^{n}\sum_{l=0}^{m}%
\frac{m!n!\left(  \sqrt{2}\right)  ^{k+l}}{\left(  n-k\right)  !k!\left(
m-l\right)  !l!}\nonumber \\
&  \times%
\genfrac{}{}{0pt}{}{:}{:}%
H_{n-k,m}\left[  \left(  Q_{1}-Q_{2}\right)  ,i\left(  P_{1}-P_{2}\right)
\right]  H_{m-l,n}\left[  i\left(  Q_{1}+Q_{2}\right)  ,-\left(  P_{1}%
+P_{2}\right)  \right]
\genfrac{}{}{0pt}{}{:}{:}%
, \label{g3}%
\end{align}
where we have used the following integration formula\cite{r13}%
\begin{equation}
\int \frac{dxdy}{\pi}x^{m}y^{r}\exp[2i\left(  y-s\right)  \left(  x-t\right)
]=\left(  \frac{1}{\sqrt{2}}\right)  ^{m+r}\left(  -i\right)  ^{r}%
H_{m,r}\left(  \sqrt{2}t,i\sqrt{2}s\right)  , \label{g4}%
\end{equation}
with $H_{m,r\text{ }}$is the two-variable Hermite polynomials \cite{r14},%
\begin{equation}
H_{m,r}(t,s)=\sum_{l=0}^{\min(m,r)}\frac{m!r!(-1)^{l}}{l!(m-l)!(r-l)!}%
t^{m-l}s^{r-l}. \label{g5}%
\end{equation}
Eq.(\ref{g3}) is a simple approach to turn $\left(  a_{1}^{\dagger}%
-a_{2}\right)  ^{n}\left(  a_{1}+a_{2}^{\dagger}\right)  ^{m}$ into its Weyl
ordering. Similarly, using Eqs.(\ref{c11}) and (\ref{g2}), we see that the
Weyl ordered form of $\left(  a_{1}+a_{2}^{\dagger}\right)  ^{m}\left(
a_{1}^{\dagger}-a_{2}\right)  ^{n}$ is
\begin{align}
&  \left(  a_{1}+a_{2}^{\dagger}\right)  ^{m}\left(  a_{1}^{\dagger}%
-a_{2}\right)  ^{n}\nonumber \\
&  =-\left(  \frac{1}{2}\right)  ^{m+n}\sum_{k=0}^{n}\sum_{l=0}^{m}%
\frac{\left(  -1\right)  ^{l}m!n!\left(  \sqrt{2}\right)  ^{k+l}}{\left(
n-k\right)  !k!\left(  m-l\right)  !l!}\nonumber \\
&  \times%
\genfrac{}{}{0pt}{}{:}{:}%
H_{n,m}\left[  \left(  Q_{2}-Q_{1}\right)  ,i\left(  P_{1}-P_{2}\right)
\right]  H_{m-k,n-l}\left[  \left(  Q_{1}+Q_{2}\right)  ,i\left(  P_{1}%
+P_{2}\right)  \right]
\genfrac{}{}{0pt}{}{:}{:}%
. \label{g6}%
\end{align}

In summary, as a natural extension of our preceding paper\cite{r7},
through employing operators' Weyl ordering expansion the formula we
have found a new two-fold complex integration transformation about
the Wigner operator $\Delta \left(  \mu,\nu \right)  $ (in its
entangled form) in phase space quantum mechanics, which is
invertible and obeys Parseval theorem. The new complex integration
transformation is compatible to $\eta-\xi$ phase space quantum
mechanics which takes entangled state representations as basis.
Under this transformation the relationship between the Weyl ordering
and $\left( a_{1}^{\dagger}-a_{2}\right)  \Leftrightarrow \left(
a_{1}+a_{2}^{\dagger }\right)  $ ordering or $\left(
a_{1}+a_{2}^{\dagger}\right)  \Leftrightarrow \left(
a_{1}^{\dagger}-a_{2}\right)  $ ordering of operators is revealed.
In this way, the contents of phase space quantum mechanics\cite{r5}
can be further enriched. Studies about the multi-mode case are in
progress and new results will be reported later.

\end{document}